\documentclass[aps,prl,twocolumn,superscriptaddress,showpacs,floatfix,altaffilletter,longbibliography]{revtex4-2}
\usepackage{graphicx}
\usepackage{grffile}
\usepackage{amsmath}
\usepackage{amssymb}
\usepackage{bm}
\usepackage{color}
\usepackage[dvipsnames]{xcolor}
\usepackage{amsmath,amssymb,amsfonts}
\usepackage{epsfig}
\usepackage{times}
\usepackage[colorlinks,bookmarks=false,citecolor=blue,linkcolor=blue,urlcolor=blue]{hyperref}
\usepackage{dsfont}
\usepackage{multirow}
\usepackage{ulem} 

\usepackage[customcolors,shade]{hf-tikz}

\usepackage{colortbl}

\newcommand{\fref}[1]{Fig.~\ref{#1}}
\newcommand{\eref}[1]{Eq.~(\ref{#1})}

\begin{document}

\title{
    Frequency-dependent Inter-pseudospin Solutions to Superconducting Strontium Ruthenate
}

\author{O.~Gingras}
\email{ogingras@flatironinstitute.org}
\affiliation{Center for Computational Quantum Physics, Flatiron Institute, 162 Fifth Avenue, New York, New York 10010, USA}
\affiliation{D\'epartement de Physique, Institut quantique, Universit\'e de Sherbrooke, Sherbrooke, Qu\'ebec, Canada}
\affiliation{D\'epartement de Physique, Universit\'e de Montr\'eal, C. P. 6128, Succursale Centre-Ville, Montr\'eal, Qu\'ebec H3C 3J7, Canada}

\author{N.~Allaglo}
\affiliation{D\'epartement de Physique, Institut quantique, Universit\'e de Sherbrooke, Sherbrooke, Qu\'ebec, Canada}

\author{R.~Nourafkan}
\affiliation{D\'epartement de Physique, Institut quantique, Universit\'e de Sherbrooke, Sherbrooke, Qu\'ebec, Canada}

\author{M.~C\^ot\'e}
\affiliation{D\'epartement de Physique, Universit\'e de Montr\'eal, C. P. 6128, Succursale Centre-Ville, Montr\'eal, Qu\'ebec H3C 3J7, Canada}

\author{A.-M.~S.~Tremblay}
\affiliation{D\'epartement de Physique, Institut quantique, Universit\'e de Sherbrooke, Sherbrooke, Qu\'ebec, Canada}

\date{\today}

\begin{abstract}
    The lasting puzzle of the superconducting order parameter of Sr$_2$RuO$_4$ calls for theoretical studies that include seldom-considered effects.
    Here we include spin-orbit coupling effects on the electronic structure and then solve the linearized Eliashberg equation in a pseudospin basis, including the possibility that spin and charge fluctuations induce frequency-dependent superconducting order parameters.
    We find that spin-orbit coupling mixes even and odd contributions in orbital, spin and frequency spaces and that leading inter-pseudospin symmetries, B$_{1g}^+$ and A$_{2g}^-$, have intra-orbital components respectively even and odd in Matsubara frequency.
    An accidental degeneracy between these could resolve apparent experimental contradictions.
\end{abstract}

\maketitle

For several decades, the paradigm of $s$-wave superconductivity has been found inadequate to describe superconductivity in correlated systems.
A large variety of superconducting order parameters (SCOP) have already been identified.
For example, cuprate high-temperature superconductors are $d$-wave with B$_{1g}$ symmetry~\cite{scalapino_case_1995}.
Pnictides can exhibit $s^\pm$-wave, an A$_{1g}$ symmetry with a sign change between different Fermi surfaces~\cite{wu_boundary-obstructed_2020}.
Usually, the symmetry of the SCOP of new systems is identified rather quickly.
It is thus surprising that after several decades of work and recent remarkable progress of experimental probes, the symmetry of the SCOP of Sr$_2$RuO$_4$ (SRO) has not yet been unambiguously established~\cite{mackenzie_even_2017, huang_review_2021}.
The reason is that certain measurements appear contradictory.
This is not only an experimental challenge but also one for theories of strong electronic correlations in multi-orbital system with important spin-orbit coupling (SOC), where a large variety of symmetries are possible~\cite{kaba_group-theoretical_2019, PhysRevB.100.134506}.

Initially reckoned a spin-triplet state due to its constant Knight-shift~\cite{ishida_spin-triplet_1998, mackenzie_superconductivity_2003}, independent verification has highlighted a heating effect so that a dominantly spin-singlet state appears more credible~\cite{pustogow_constraints_2019, ishida_reduction_2020, chronister_evidence_2021}.
Another experiment probing spins using polarized neutrons met a similar fate~\cite{duffy_polarized-neutron_2000, petsch_reduction_2020}.
Previously in contradiction with evidences for Pauli limiting~\cite{mackenzie_even_2017, steppke_strong_2017}, these experiments now agree. 

Another critical characteristic of SRO is its two-component nature inferred by evidences of time-reversal (TR) symmetry breaking~\cite{luke_time-reversal_1998, xia_high_2006}.
Ultrasound experiments also support a two-component SCOP which couples to the B$_{2g}$ shear mode~\cite{benhabib_ultrasound_2021, ghosh_thermodynamic_2021}.
Additionally, the enhancement of the critical temperature under uniaxial pressure~\cite{steppke_strong_2017} not only provides a strong evidence for an even-parity (\textit{e-p}) SCOP~\cite{sunko_direct_2019}, it is also a useful knob to study this two-component property. 
Indeed, muon spin relaxation ($\mu$SR) measurements observed two transition temperatures under pressure, indicative of a lift in degeneracy between the two components~\cite{grinenko_split_2021}.
But specific heat measurements, extremely sensitive to superconducting transitions, detected a single transition temperature~\cite{li_high-sensitivity_2021}.

Consequently, various theoretical proposals have been formulated in replacement to the initial chiral $p$-wave~\cite{Kallin_2009}, including domain-wall physics and inhomogeneities~\cite{ghosh2021strong, Schmalian_Inhomogeneous_2021}.
A two-component character can be realized in two ways.
First, the components can be degenerate by symmetry if the SCOP transforms like a two-dimensional (2D) irreducible representation (irrep) of the $D_{4h}$ point group.
The only \textit{e-p} such possibility is the E$_g$ irrep.
One such proposed state is the $d_{xz}+id_{yz}$ which could originate from momentum dependent $\textbf{k}$-SOC~\cite{suh_stabilizing_2020, clepkens_shadowed_2021}.
However, density-functional theory (DFT) expects this coupling to be negligibly small in SRO, known to have a quasi-2D character~\cite{hussey_normal-state_1998, PhysRevLett.101.026406}.
Moreover, these E$_g$ states under uniaxial stress should generate two transitions in specific heat.
Another symmetry protected possibility is an odd-orbital spin-singlet odd-frequency (odd-$\omega$) state~\cite{kaser_inter-orbital_2021}, which is gapless contrary to experiments.

Another possibility for two components is that they are degenerate by accident and transform like different irreps.
The most natural of the two components is a $d_{x^2-y^2}$ B$_{1g}$ state, since thermal conductivity and scanning tunneling microscopy point in this direction~\cite{hassinger_vertical_2017, sharma_momentum-resolved_2020} and it should originate from antiferromagnetic fluctuations predicted by DFT~\cite{gingras_superconducting_2019}.
Such a symmetry was well studied in the context of the cuprates~\cite{scalapino_case_1995}.
For the second component, some works have proposed an extended $s$-wave~\cite{raghu_theory_2013, romer_knight_2019, romer_fluctuation-driven_2020} or odd-parity (\textit{o-p}) states~\cite{PhysRevB.89.220510, PhysRevLett.121.157002, PhysRevResearch.1.033108,  scaffidi_degeneracy_2020, PhysRevResearch.3.L042002} originating from spin fluctuations caused by the nesting of the quasi-one-dimensional bands~\cite{sidis_evidence_1999}.
Unfortunately, these combination would not couple to the B$_{2g}$ shear mode.
Other works proposed $g_{xy(x^2-y^2)}$ A$_{2g}$, a higher angular momentum version of $d_{x^2-y^2}$~\cite{kivelson_proposal_2020, yuan_strain-induced_2021, clepkens_higher_2021, Wagner_GL_2021}.
The similar nodal structures of $d_{x^2-y^2}$ and $g_{xy(x^2-y^2)}$ could reduce the signature on specific heat, but not remove it entirely.
It has been proposed theoretically that this accidental degeneracy is more consistent with ultrasound experiments than the other symmetry-protected $d_{xz}+id_{yz}$ proposal~\cite{sheng2021superconducting}. 
Moving away from the $d_{x^2-y^2}$ state, a $d_{xy}\pm i s^*$ state was proposed~\cite{romer_superconducting_2021,bhattacharyya2021superconducting}.

Accidental degeneracies should be lifted by small perturbations.
Although not definitive, $\mu$SR measurements under isotropic conditions did not observe a split in the critical temperatures~\cite{grinenko_unsplit_2021}.
Additionally, disorder by non-magnetic impurities could help split the transition temperatures~\cite{zinkl_impurity-induced_2021}.

\begin{figure*}
    \includegraphics[width=.82\linewidth]{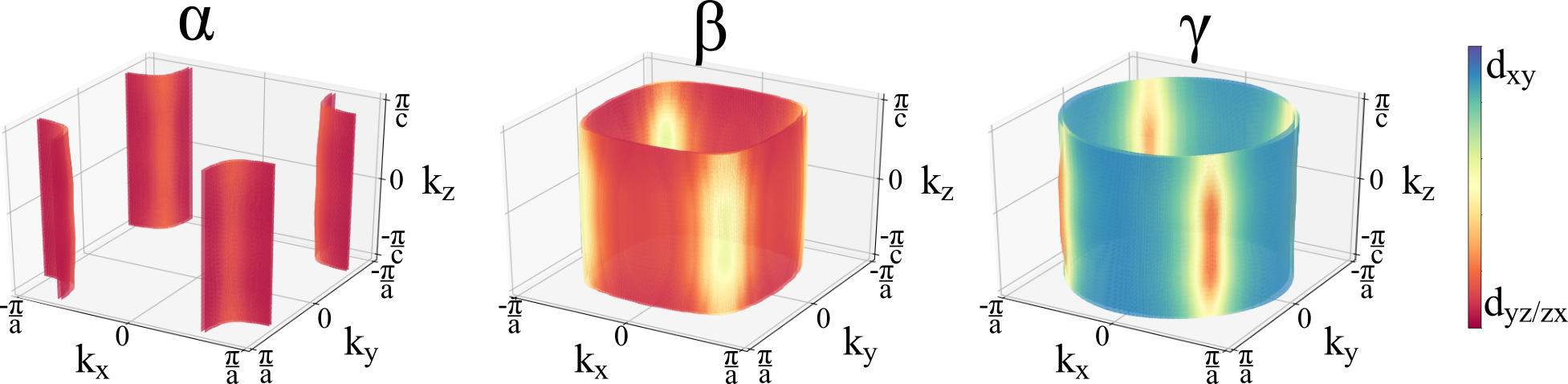}
    \caption{\label{fig:orb_char}
    Orbital character projected on the $\alpha$, $\beta$ and $\gamma$ Fermi sheets as also calculated in Ref.~\citenum{PhysRevLett.112.127002}. 
    The blue color corresponds to $xy$ orbital character while the red color corresponds to $yz$ and $xz$ orbital characters without distinction between these two. }
\end{figure*}

The broad richness of possible superconducting states in SRO is due to the extreme sensitivity of pairing interactions to the normal state electronic structure and many-body effects.
Most theoretical studies have so far neglected SOC whose importance for SRO has been demonstrated in detailed comparisons between photoemission spectroscopy and first-principles calculations~\cite{PhysRevLett.101.026406, PhysRevLett.112.127002, tamai_high-resolution_2019}.
Here, it is included on the normal state electronic structure that we project on the relevant $t_{2g}$ subspace.
Spin and charge fluctuations driven by electronic correlations in that subspace provide the pairing interaction~\cite{Bourbonnais_1986,Scalapino_Hirsch_1986,scalapino_common_2012}.
We treat their frequency-dependence and find that SOC leads to coexistence between even- and odd-$\omega$ components.
Two leading superconducting states dominates: the usual B$_{1g}^+$ state and an A$_{2g}^-$.
Their intra-orbital (intra-$l$) components are respectively even- and odd-$\omega$.
Their accidental degeneracy could resolve the observed contradictions.

\paragraph{Normal state. --} 
The layered perovskite structure of SRO is characterized by the $D_{4h}$ space group.
The main physics resides within the ruthenium (Ru) and oxygen (O) planes, making it quasi-2D.
The Ru atom is at the center of an elongated octahedron of O atoms.
The resulting crystal field splits its $4d$ electrons into the unoccupied $e_g$ and the partially filled $t_{2g}$ subsets.
The $t_{2g}$ orbitals $xy$, $yz, xz$, hybridize with the O's $p$ electrons to form the $\alpha$, $\beta$ and $\gamma$ Fermi sheets shown in \fref{fig:orb_char}.
Since the $e_g$ orbitals are far from the Fermi level, the electronic fluctuations responsible for mediating superconductivity are be solely hosted by the $t_{2g}$ orbitals. 

We compute the electronic structure from the projected-augmented-wave pseudopotential~\cite{blochl_projector_1994, amadon_plane-wave_2008} ABINIT implementation~\cite{gonze_abinit_2020, romero_abinit_2020} of DFT in the local density approximation~\cite{hohenberg_inhomogeneous_1964, kohn_self-consistent_1965}.
We downfold the bands onto the $t_{2g}$ states.
The large SOC on the Ru atom couples the $xy$ with the $xz$ and $yz$ orbitals with opposite spins.
Without SOC, spin conservation makes the normal state Hamiltonian diagonal in spins while crystal-field symmetry preserves the block diagonal form of the $xz,yz$ sector.
With SOC, the spin and orbital sectors are coupled~\cite{PhysRevLett.112.127002}.
The colors in \fref{fig:orb_char} represent the orbital characters of the resulting Fermi sheets in the first Brillouin zone.
The $\alpha$ and $\beta$ sheets mainly comes from quasi-one-dimensional bands with $xz$ and $yz$ orbital characters, while the $\gamma$ sheet is mostly a quasi-2D band with $xy$ orbital character.
The color code clearly shows that SOC introduces spin-orbital entanglement around the k$_x = \pm \text{k}_y$ diagonals.

Considering local SOC, the Hamiltonian can be block diagonalized into pseudospin up ($+$) and down ($-$) sectors denoted by $\rho=\pm$.
Non-local $\textbf{k}$-dependent SOC effects arise from SOC on the O atoms when the Hamiltonian is downfolded onto $t_{2g}$ orbitals. 
Through inter-layer coupling, this mechanism breaks pseudospin symmetry.
However, this effect is negligibly small at the DFT level~\cite{clepkens_shadowed_2021} and is less enhanced by correlations than local SOC~\cite{SOC_Local_non-local}.
We can thus work in a 2D $\textbf{k}$-space where normal state Green functions $\bm G_{Kl_1l_2}^{\rho}$ are block diagonal in the pseudospin $\rho$ and energy-momentum $K \equiv (i\omega_m, \textbf{k})$ basis but need an orbital $l_1$ to $l_2$ dependence. 

\paragraph*{Linearized Eliashberg equation. --}
Details and derivations are given in the companion paper~\cite{gingras_PRB_2022}.
The pairing mechanism is contained in an effective interaction that combines with the pair susceptibility to yield a pairing function $\bm V_{pp}$ that must be diagonalized to find the superconducting instabilities of the normal state.
This can be written in the form of a linearized Eliashberg equation~\cite{esirgen_fluctuation_1998,bickers_self-consistent_2004} 
 \begin{equation}
    \label{eq:eliashberg}
    \lambda \bm \Delta(\bm 1, \bm 2) = - \bm V_{pp}(\bm 1, \bm 2, \bar{\bm 1}, \bar{\bm 2}) \bm \Delta(\bar{\bm 1}, \bar{\bm 2}).
\end{equation}
Boldface numerals such as $\bm 1$ stand for Matsubara frequencies and quantum numbers described above. 
Overbars stand for implied integration/summation.
Given the linear nature of the equation, the eigenvectors $\bm \Delta$ transform as the irreps of $D_{4h}$ and a superconducting instability occurs when one of the eigenvalues $\lambda$ reaches unity.
The corresponding $\bm \Delta$ reflects the symmetry of the Gorkov function $\langle T_{\tau} \psi(\bm 1) \psi(\bm 2) \rangle$ just below the transition temperature.
Here, $\psi(\bm 1)$ is the destruction operator and $T_{\tau}$ is the imaginary-time-ordering operator.

Isoelectronic doping suggests that SRO lies in the vicinity of magnetic orderings~\cite{carlo_new_2012}, consistent with the important spin fluctuations found by neutron scattering~\cite{sidis_evidence_1999,iida_inelastic_2011,steffens_spin_2019, jenni_neutron_2021}. 
In addition, the well established correlated character of Ru $t_{2g}$ electrons~\cite{mravlje_coherence-incoherence_2011, Rozbicki_2011, zhang_fermi_2016,kim_spin-orbit_2018, strand_magnetic_2019, kugler_strongly_2020} makes SRO the archetypal representative of superconductivity mediated by spin and charge fluctuations.

\paragraph{Spin and charge fluctuations. --}
The bare interactions between electrons are diagonal in the basis of isolated atoms.
We model these with the rotationally invariant Kanamori-Slater Hamiltonian (KSH)~\cite{georges_strong_2013} characterized by two parameters: the on-site Coulomb repulsion $U$ and Hund's coupling $J$.
Since $J$ favours same spin alignment, the inter-orbital (inter-$l$) repulsion is stronger in the inter-spin channel $U'= U-2J$ than in the intra-spin one $U''=U-3J$, imposing $J/U \geq 0$.

In the paring function $\bm V_{pp}$, the binding glue for electron pairs consists of low-energy collective modes formed by an avalanche of electron-hole pairs.
In other words, electrons scatter-off each other through exchange of spin- and charge-density fluctuations.
These fluctuations are captured by the particle-hole (\textit{p-h}) polarizability $\bm \chi_{ph}$.
Here, we compute $\bm \chi_{ph}$ using the random phase approximation (RPA) where the irreducible \textit{p-h} vertex $\bm \Gamma_{ph}$ is replaced by local interactions that have the symmetries of the local interactions described above.
The resulting phase diagrams exhibit a rich variety of competing ordered states in either the \textit{p-h} or the particle-particle (\textit{p-p}) channels.

The interaction-induced enhancement of spin (charge) fluctuations is quantified by the magnetic (density) Stoner factor $S^{m(d)}$.
We enforce $S^{m(d)} < 1$ to prevent the system from falling into a magnetic (charge) ordered state.
In spin- and charge-fluctuation mediated superconductivity, $\bm \chi_{ph}$ serves as an effective interaction for pairing electrons in the \textit{p-p} channel.
The irreducible pairing vertex $\bm \Gamma_{pp}$ is expressed in terms of $\bm \chi_{ph}$ and $\bm \Gamma_{ph}$ through Parquet-like equations, but without the self-consistency~\cite{bickers_self-consistent_2004}.
The full pairing susceptibility $\bm \chi_{pp}(Q)$ is thus enhanced compared to the bare \textit{p-p} susceptibility $\bm \chi_{pp}^0(Q)$. 
In the absence of magnetic fields, an instability cascades into a superconducting phase when $\bm \chi_{pp}(Q=0)$ diverges, or equivalently when the pairing function $\bm V_{pp} \equiv \bm \Gamma_{pp}(0)\bm \chi^0_{pp}(0)$ has an eigenvalue equal to unity, as seen in \eref{eq:eliashberg}.
We are interested in the symmetry of the SCOP $\bm \Delta$.

\paragraph*{Pseudospin basis. --}
The KSH conserves spins but not pseudospins~\cite{gingras_PRB_2022,romer_knight_2019}.
However, it is still block diagonal
and $\bm V_{pp}$ can be decomposed into the intra- and inter-pseudospin (intra-$\rho$ and inter-$\rho$) channels.
Inversion symmetry forces all solutions to have \textit{e-p} or \textit{o-p} and
Pauli principle leads to
$
    \bm \Delta^{ep/op}_{Kl_1l_2} = \bm \Delta^{+-}_{Kl_1l_2} \mp \bm \Delta^{-+}_{K^*l_2l_1}
$
where $K^* \equiv (-i\omega_m, \textbf{k})$. In this basis, \eref{eq:eliashberg} can be reduced to 
\begin{equation}
    \label{eq:effective_eliashberg}
    \lambda \bm \Delta_K^{\alpha}
    =
    -\frac{1}{2}
    \sum_{K'}
    \bm V^{\alpha}_{KK'}
    \bm \Delta_{K'}^{\alpha}
\end{equation}
for $\alpha$ the \textit{e-p} (\textit{o-p}) channel, with the effective vertices $\bm V^{\alpha}_{KK'}$ given in Ref.~\citenum{gingras_PRB_2022}.
This problem does not introduce new numerical challenges compared with the spin-diagonal case.

The intra-$\rho$ channel however could not be simplified using Pauli principle. 
Fortunately, we observe that intra-$\rho$ solutions are nearly degenerate to an inter-$\rho$ one, thus the inter-$\rho$ channel is sufficient to discuss solutions.
Moreover, we note that \eref{eq:effective_eliashberg} applies regardless of the method to compute $\bm \Gamma_{ph}$ as long as it satisfies the pseudospin block diagonal property.

\paragraph{Results. --}
In principle, the parameters $U$ and $J$ entering the KSH can be calculated, but their effective values are different due to screening.
We thus explore this parameter space. 
Instead of $U$ and $J$, we quantify the interactions by $J/U$ and $S^m$.
In addition to $J>0$, the repulsive nature of the KSH constrains $J/U < 1/3$ but we consider $J/U \leq 0.45$ to allow some attractive interactions. 
$S^m$ quantifies the proximity to a magnetic transition.
The closer to  $S^m=1$, the larger the magnetic fluctuations.
Because SRO is in proximity to magnetic orderings~\cite{carlo_new_2012}, we study $S^m \geq 0.5$ to tune the system in the vicinity of a magnetic instability.
We never reach $S^m \ (S^d) \geq 1$ since it would correspond to a magnetic (charge) instability of the \textit{p-h} channel.

\begin{figure}[t!]
    \centering
    \includegraphics[width=.79\linewidth]{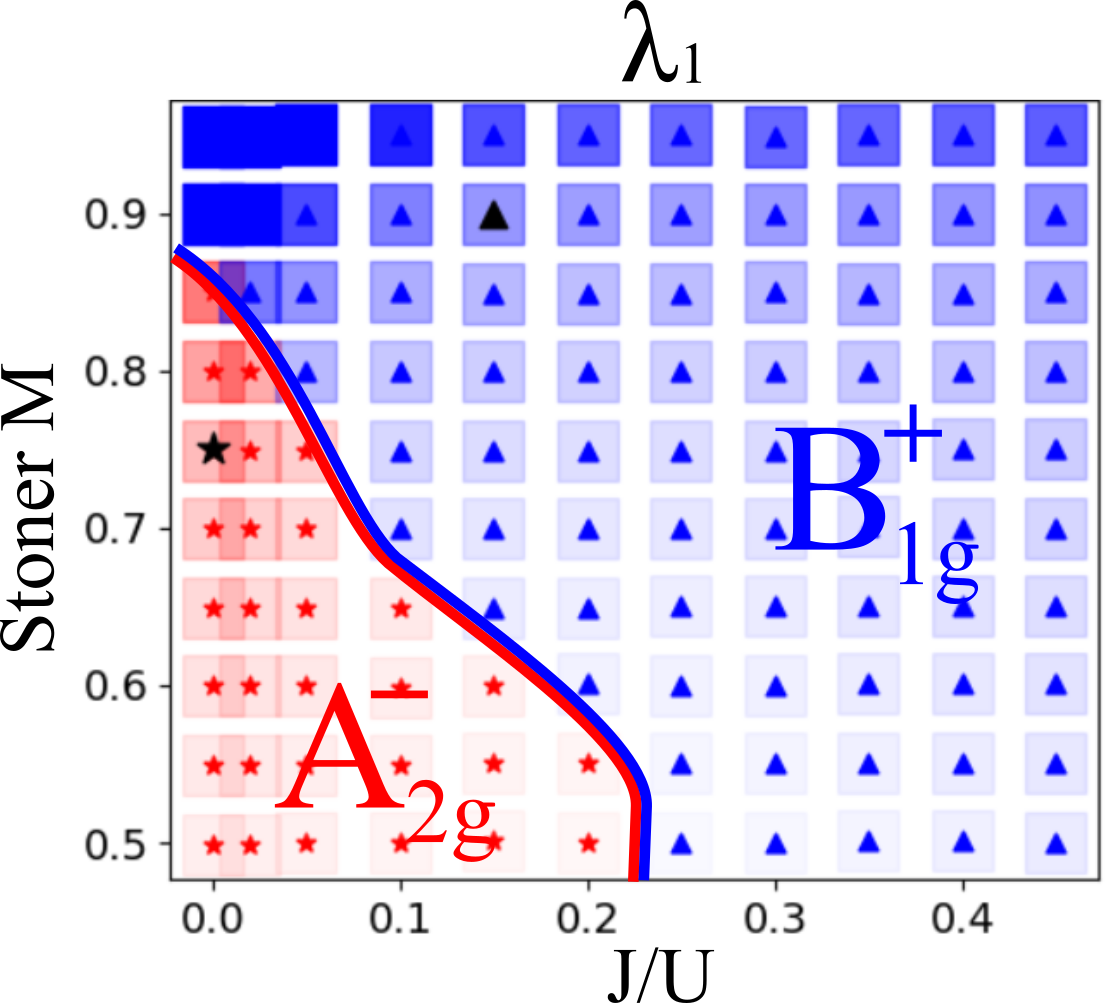}
    \caption{(Color online) Global irrep of \eref{eq:effective_eliashberg}'s leading eigenvectors at $T=250$~K. 
    A square's transparency represents the eigenvalue between 0 and 1. Notice a bigger black triangle $\blacktriangle$ (star $\bigstar$) symbol for $J/U = 0.15$ ($0$) and $S^m=0.9$ ($0.75$), referred to in the text.}
    \label{fig:phase_diag_eig1}
\end{figure}

\begin{figure*}
    \centering
    \includegraphics[width=.95\linewidth]{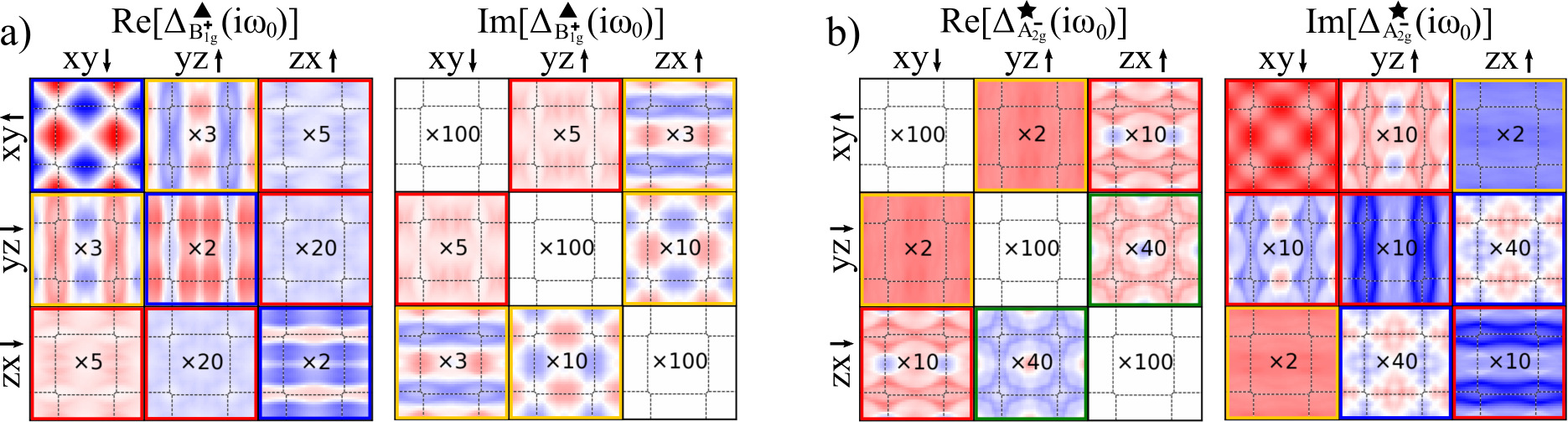}
    \caption{Real and imaginary parts at $i\omega_0$ of the leading gap functions denoted by the a) $\blacktriangle$ (B$_{1g}^+$) and b) $\bigstar$ (A$_{2g}^-$) symbols in \fref{fig:phase_diag_eig1}. Each $3\times 3$ matrix is for one pseudospin component. Each square in the matrix is the momentum distribution of a component of the gap function. It is shown in the first Brillouin zone for k$_z=0$ and k$_z=2\pi/c$. The color goes from $-1$ (blue) to $1$ (red) and components were rescaled with a coefficient printed at their respective $\Gamma$ point. The colors around the squares show the $SPOT$ character of each component as in Table~\ref{tab:spot_decomposition_b1g_a2g}.}
    \label{fig:gaps_b1g_a2g}
\end{figure*}

\fref{fig:phase_diag_eig1} shows a phase diagram in the $S^m$ \textit{vs} $J/U$ parameter space of the irrep that characterizes how the leading eigenvector of \eref{eq:effective_eliashberg} transforms. 
An irrep classifies how symmetries of the normal state are broken upon entering the ordered state.
For SCOPs, $\bm \Delta_{K}^{\mu_1\mu_2}$ transforms like two electrons with four quantum numbers each: the energy-momenta $K$ and $-K$ and the spin-orbital indices $\mu_1\equiv(\sigma_1, l_1)$ and $\mu_2$.
The normal state transforms spatially like the $D_{4h}$ group.
It also has TR symmetry, which is always satisfied within a global phase.
Given that even- and odd-$\omega$ contributions can mixed when pairing between different orbitals is considered~\cite{black-schaffer_odd-frequency_2013}, we label the irreps with $+/-$~\cite{geilhufe_symmetry_2018,gingras_PRB_2022} based on the intra-$l$ components since these are pure in even-/odd-$\omega$.

Another way to characterize SCOPs is to use the $SPOT$ classification~\cite{linder_odd-frequency_2019} or generalizations~\cite{PhysRevResearch.3.033255}.
The operations of exchanging the spins, momenta, orbitals and frequencies are respectively designated by $\hat{S}$, $\hat{P}^*$, $\hat{O}$ and $\hat{T}^*$.
Using those, the Pauli principle for a two-fermion gap function reads $\hat{S}\hat{P}^*\hat{O}\hat{T}^*\bm \Delta = - \bm \Delta$.
They are idempotent, thus each operation has eigenvalue $\pm 1$ labelled $S,P,O,T$.
Since $D_{4h}$ has inversion symmetry, $\hat{P}^*\bm \Delta = + (-) \bm \Delta$ and corresponding irreps are labelled by $g$ ($u$), respectively noted $^+P$ ($^-P$).
Neglecting SOC, we similarly have $\hat{S}\bm \Delta = + (-) \bm \Delta$ for triplet $^+S$ (singlet $^-S$) solutions.
However, SOC in multi-orbital systems introduces spin-flip and inter-$l$ interactions that mix $^+S$ and $^-S$ solutions~\cite{PhysRevLett.112.127002}, along with $^{\pm}O$ and $^{\pm}T$.
Consequently SCOPs no longer are eigenvectors of either $\hat{S}$, $\hat{O}$ and $\hat{T}^*$ operators.

\begin{table}[ht!]
    \centering
    \caption{$SPOT$ decompositions of the gap functions in Figs~\ref{fig:gaps_b1g_a2g}. Here, $\mathcal{P}^{SPOT}\bm \Delta$ is the ratio of the absolute values of the projected gap function $\bm \Delta$ for a specific $SPOT$ on the total one.}
\begin{tabular}{c|c|c|c|c|c}
    \label{tab:spot_decomposition_b1g_a2g}
    \centering
    \ $S$ \quad & \ $P$ \quad & \ $O$ \quad & \ $T$ \quad & $\mathcal{P}^{SPOT} \bm \Delta^{\blacktriangle}({\text{B}_{1g}^+})$ & $\mathcal{P}^{SPOT} \bm \Delta^{\bigstar}({\text{A}_{2g}^-})$  \\
    \hline
    \hline
    \rowcolor{blue!30} - & + & + & + & 95\% & $<$1\% \\
    \rowcolor{red!30} + & + & + & - & 10\% & 81\% \\
    \rowcolor{orange!30} + & + & - & + & 29\% & 58\% \\
    \rowcolor{Green!30} - & + & - & - &     0\% & $<$1\%
\end{tabular}
\end{table}

Only two different irreps appear in the phase diagram \fref{fig:phase_diag_eig1}: a B$_{1g}^+$ and a A$_{2g}^-$.
For each, we selected a SCOP labelled by $\blacktriangle$ and $\bigstar$ respectively.
In Table~\ref{tab:spot_decomposition_b1g_a2g}, we show their respective projections on $SPOT$ eigenvectors.
In \fref{fig:gaps_b1g_a2g}, we display the real and imaginary momentum distribution in the pseudospin-orbital basis at $i\omega_0$ of $\blacktriangle$ in a) and $\bigstar$ in b).

We first discuss the $\blacktriangle$ (B$^+_{1g}$) SCOP.
Details are given in Ref.~\citenum{gingras_PRB_2022}.
All the contributions are \textit{e-p} $^+P$.
The intra-$l$ pairs form spin-singlet $^-S$ with even-$\omega$ $^+T$ character.
We fix the global phase such that they are purely real.
The $l_1 = l_2 = xy$ component is the largest and, as seen in the $11$ component in \fref{fig:gaps_b1g_a2g}~a), it transforms like a B$_{1g}$ $d_{x^2-y^2}$ function in $\textbf{k}$-space.
It was attributed to the $\gamma$ band's antiferromagnetic nesting vector~\cite{gingras_superconducting_2019}.
Globally, these pairs transform like B$_{1g}^+$ with $^-S^+P^+O^+T$ character.

Because one of the paired $xy$ electrons can propagate to the $\{xz,yz\}$ orbitals by flipping a spin yet preserving pseudospin, SOC generates inter-$l$ pairs which now have equal spins, {\it i.e.} $^+S$.
These $xy;\{yz, zx\}$ inter-$l$ components are comparable in magnitude to the $xy;xy$ one.
They transform globally like B$_{1g}$.
There are two possibles $SPOT$ characters that can entangle without affecting TR symmetry: $^+S^+P^-O^+T$ and $^+S^+P^+O^-T$.
In \fref{fig:gaps_b1g_a2g}~a), the real (imaginary) parts of 
$\bm \Delta_{yz,xy}^{\downarrow\downarrow}$ and $\bm \Delta_{xy, yz}^{\uparrow\uparrow}$ ($\bm \Delta_{zx, xy}^{\downarrow\downarrow}$ and $\bm \Delta_{xy, zx}^{\uparrow\uparrow}$) have $^+S^+P^-O^+T$ character while their complementary imaginary (real) parts have $^+S^+P^+O^-T$ character.

Finally, the other $xy$ electron can also flip its spin and propagate to $\{yz, zx\}$ orbitals.
All orbital sectors are connected and the intra-$l$ $\bm \Delta_{yz,yz}^{-\sigma,\sigma}$ and $\bm \Delta_{zx, zx}^{-\sigma,\sigma}$ are the second largest components.
They globally transform like the $\bm \Delta_{xy, xy}^{\sigma, -\sigma}$ component as B$_{1g}^+$ with $^-S^+P^+O^+T$ character.

This B$_{1g}^+$ state is a prime candidate for accidental degeneracies in SRO, yet the entanglement of $SPOT$ characters was rarely discussed.
We see that SOC couples all orbital sectors into a single global irrep with ubiquitous even- and odd-$\omega$ correlations~\cite{black-schaffer_odd-frequency_2013}.
Although arguably responsible for the finite polar Kerr effect~\cite{komendova_odd-frequency_2017}, they are insuffisant to explain the two-component signatures in SRO, which motivates revisiting the polar Kerr experiment under uniaxial pressure. 

Moving to the $\bigstar$ (A$^-_{2g}$) SCOP shown in \fref{fig:gaps_b1g_a2g}~b), we fix the global phase so the $^-T$ intra-$l$ components are purely imaginary.
They transform like $^+S^+P^+O^-T$.
Because they form $^+S$ spin-triplets for $\sigma_1 = - \sigma_2$, a spin-flip induced by SOC preserves $^+S$ in inter-$l$ sectors.
Those have dominant $^+S^+P^-O^+T$ and subdominant $^+S^+P^+O^-T$ characters.

The interest in this SCOP is two-fold.
First, it transforms like A$_{2g}$, consistent with ultrasound experiments.
Second, intra-$l$ odd-$\omega$ SCOP are gapless at the Fermi surface and their contribution to the specific heat could be extremely subtle.
Consequently, an accidental degeneracy between B$_{1g}^+$ and A$_{2g}^-$ could explain the discrepancy between specific heat and $\mu$SR under uniaxial strain, motivating further investigations.

Finally, the absence of the $d_{yz}\pm id_{zx}$ (E$_g^+$)~\cite{suh_stabilizing_2020, clepkens_shadowed_2021},  E$_g^-$~\cite{kaser_inter-orbital_2021}, $s^*$-wave (A$_{1g}^+$)~\cite{raghu_theory_2013, romer_knight_2019, romer_fluctuation-driven_2020} and $g_{xy(x^2-y^2)}$ (A$_{2g}^+$)~\cite{kivelson_proposal_2020, yuan_strain-induced_2021, clepkens_higher_2021, Wagner_GL_2021} solutions as dominant instabilities is explained by the requirement from SOC in 2D that all orbital components participate in pairing and transform like a unique global irrep.
Instead, we find that most subleading eigenvectors have odd-$\omega$ intra-$l$ components~\cite{gingras_PRB_2022}.

\paragraph{Summary. --}
We found that SOC leads to multi-orbital SCOP with multiple $SPOT$ contributions.
In proximity to a magnetic instability, we found the B$_{1g}^+$ state often considered to explain SRO. 
Further away from the magnetic instability, we found an A$_{2g}^-$ whose intra-$l$ components are odd-$\omega$.
Their combination into an accidental degeneracy could explain superconducting SRO since
this solution has the potential to explain all experiments, notably the discrepancy between specific heat and $\mu$SR for uniaxial strain.
Their study requires calculations of observables for gap functions that contain odd-$\omega$ contributions. 

\begin{acknowledgments}
    \textit{Acknowledgments. --}
    We are grateful for discussions with Michel Ferrero, Sékou-Oumar Kaba and David Sénéchal.
    This work has been supported by the Fonds de Recherche du Qu\'ebec—Nature et Technologie (FRQNT), the Hydro-Qu\'ebec fellowship, and the Université de Montréal (OG), the Research Chair in the Theory of Quantum Materials, the Canada First Research Excellence Fund, the Natural Sciences and Engineering Research Council of Canada (NSERC) under Grants No. RGPIN-2014-04584, No. RGPIN-2019-05312 (AMST), and No. RGPIN-2016-06666 (MC).
    Simulations were performed on computers provided by the Canada Foundation for Innovation, the Minist\`ere de l'\'Education, du Loisir et du Sport (MELS) (Qu\'ebec), Calcul Qu\'ebec, and Compute Canada. 
    The Flatiron Institute is a division of the Simons Foundation.
    The authors are members of the Regroupement qu\'ebécois sur les matériaux de pointe (RQMP).
\end{acknowledgments}

\end{document}